\documentclass[nofootinbib,notitlepage,superscriptaddress,10pt,aps,pra,twocolumn]{revtex4-1}

\usepackage[utf8x]{inputenc}
\usepackage{amsmath}
\usepackage{amssymb}
\usepackage{graphicx}

\begin{document}

\title{Planck-scale-modified dispersion relations in FRW spacetime}
\author{Giacomo ROSATI}
\affiliation{Institute
for Theoretical Physics, University of Wroc\l{}aw, Pl.\ Maksa Borna
9, Pl--50-204 Wroc\l{}aw, Poland}

\author{Giovanni AMELINO-CAMELIA}
\affiliation{Dipartimento di Fisica, Universit\`a di Roma ``La Sapienza", P.le A. Moro 2, 00185 Roma, Italy}
\affiliation{INFN, Sez.~Roma1, P.le A. Moro 2, 00185 Roma, Italy}

\author{Antonino MARCIAN\`O}
        \affiliation{Department of Physics, Fudan University, 220 Handan Road, 200433 Shanghai, China}

\author{Marco MATASSA}
\affiliation{Department of Mathematics, University of Oslo, P.B. 1053 Blindern, 0316 Oslo, Norway}

\begin{abstract}
\noindent In recent years Planck-scale modifications of the dispersion relation
have been attracting increasing interest also from the viewpoint of possible applications
in astrophysics and cosmology, where spacetime curvature cannot be neglected. Nonetheless the interplay between Planck-scale
effects and spacetime curvature is still poorly understood, particularly in cases
where curvature is not constant. These challenges have been so far postponed by relying on an
 {\it ansatz}, first introduced by Jacob and Piran. We here
 propose a general strategy of analysis of the effects of modifications of dispersion relation in FRW spacetimes,
 applicable both to cases where
 the relativistic equivalence
of frames  is spoiled (``preferred-frame scenarios") and to the alternative possibility of ``DSR-relativistic theories", theories that are
fully relativistic but with relativistic laws deformed so that the modified dispersion relation is observer independent.
We show that the Jacob-Piran {\it ansatz}
implicitly assumes that spacetime translations are not affected by the Planck-scale, while
under rather general conditions the same Planck-scale quantum-spacetime structures producing modifications
of the dispersion relation also affect translations.
Through the explicit analysis of one of the effects produced by modifications of the dispersion relation,
an effect amounting to Planck-scale corrections to travel times, we show that our concerns are not merely
conceptual but rather can have significant quantitative implications.
\end{abstract}

\maketitle

\section{Introduction}
The possibility that Planck-scale effects might modify the ``dispersion relation",
the on-shell requirement linking energy and momentum of a particle,
has attracted quite some interest in the recent quantum-gravity literature
(see, {\it e.g.}, Refs.~\cite{gacLRR,mattinglyLRR} and references therein).
Motivation for the study of this possibility also comes from the fact that
it provides a rare case of conjectured Planck-scale feature that
could produce effects observable already with presently-available experimental technologies.
These opportunities for testing are not for controlled Earth-bound laboratory setups,
where the effects would still be too small, but rather they arise in some contexts of
astrophysics and cosmology where the ultralarge propagation distances act as amplifier
of the minute Planck-scale effects~\cite{gacLRR,mattinglyLRR}.

These opportunities from astrophysics and cosmology however also bring about
a challenge for theories, since in the relevant quantum-spacetime pictures very little
has been so far understood about the interplay between curvature and Planck-scale effects.
This is particularly true for cases where the analysis does not allow schematization
in terms of a constant spacetime curvature, as it is indeed the case when tests are
performed exploiting cosmological distances.

These challenges have been so far mostly postponed, assuming the applicability
of an {\it ansatz}, first formulated by Jacob and Piran \cite{jacobpiran} (also see Refs.\cite{Ellis:2005wr,PiranMartinez})
for this interplay between Planck-scale modifications of the dispersion relation
and spacetime curvature.

We here
 propose a general strategy of analysis of the effects of modifications of dispersion relation
 applicable when a non-constant curvature of spacetime is to be taken into account.
 We adopt a phenomenological approach: rather than attempting to establish a specific form
 of interplay between Planck-scale effects and spacetime curvature
 within one or another quantum-spacetime picture,
 we use what is presently known about the various possibilities for formalizing a
 quantum spacetime as guidance for modeling in a very general way
  the possible forms
 of this interplay.

 Of particular interest is the fact that we allow for Planck-scale features to be present not
 only in the dispersion relation but also in the description of translation transformations.
 This is important since in some of the most studied quantum spacetimes, such as the ``$\kappa$-Minkowski
 noncommutative spacetime~\cite{majidruegg,lukieKAPPAmink,bobKAPPAplb},
 dedicated analyses have shown that  the Planck scale does have this double role, affecting both the dispersion relation
 and translations.

 In this respect we uncover a particularly significant characterization of the Jacob-Piran {\it ansatz}:
 we show that this {\it ansatz} implicitly assumes that translations are unaffected by Planck-scale structures. By this we mean that in particular specializing the Jacob-Piran {\it ansatz} to the case
 of constant rate of spacetime expansion (de Sitter spacetime) one gets a picture that is invariant
 under ordinarily classical space and time translations.
 Analyses depicted as full explorations of the implications of modified dispersion relations,
which instead rely exclusively on the Jacob-Piran {\it ansatz},
 should be more carefully described as testing the dispersion relation under the
 restrictive requirement
  that translations should be unaffected by Planck-scale features.

 Another qualifying aspect of our analysis is that it considers both the
 case of modifications of the dispersion relation that signal ``LIV" (Lorentz Invariance Violation,
 {\it i.e.} preferred-frame scenarios) and the very different case where
  the modified dispersion relation is implemented  within a fully relativistic
  picture. This latter possibility of course requires deforming the relativistic laws of transformation
  among observers in a way suitable for enforcing
  the modified dispersion relation as an observer-independent law,
  a possibility that can be formalized in terms of
  the relativistic theories of type ``DSR"
(doubly-special, or, for some authors, deformed-special,
relativity), first introduced in Ref.~\cite{gacdsr1IJMPD}
(also see Refs.~\cite{jurekDSRfirst,joaoleePRDdsr}).

As well established in the recent literature, the DSR case imposes taking into account
of the novel relativistic effect of ``relative locality"~\cite{principle} (also see Refs.\cite{bob,bobKAPPAplb}). Within the scopes
of our analysis this is tedious but straightforward. However, it should be noted that we here report
the most advanced phenomenological analysis to date of the interplay between relative locality
and (non-constant) spacetime curvature.

Most of the issues we are concerned with are already present in the constant-curvature case,
as we shall show explicitly. But extending the analysis to cases with non-constant curvature
leads to encounter additional challenges. The constant-curvature case allows us to analyze
most aspects of the problem at hand in terms of pure symmetry considerations. This is
not directly available in cases with non-constant curvature; however, we recover a role for symmetries also
in the case of non-constant curvature by using the fact that locally the symmetries
of the constant-curvature case reemerge. One can in particular describe a finite path within
a FRW-type spacetime by gluing infinitesimal paths within a suitable series of deSitter-type
spacetimes, a strategy of analysis we refer to as ``thick slicing".

While the conceptual significance of our concerns should become clear to our readers very
early in the analysis here reported, it is important for us to show that our concerns
can also have significant quantitative implications. It is for this reason that we devote
much effort to the illustrative application provided by the description of the path of
a particle in a ``quantum-FRW spacetime" and the evaluation of travel times from a given source
to a given detector.

In light of the conceptual and quantitative complexity of the issues we are dealing
with we opt for focusing on the case of a 1+1-dimensional spacetime
and obtaining results only at leading order in the ultrasmall Planck length.

\section{Preliminaries on classical de Sitter spacetime and translations in a quantum spacetime}

Before starting with the main part of our analysis we find convenient to collect in this section
some known facts that we shall then use.
In the first part of this section we shall review some known facts about classical de Sitter spacetime,
the ones most relevant for our later discussion of propagation of particles in a quantum  spacetime with curvature.
In the second part we remind our readers about the interconnection found in some much-studied
quantum spacetimes between Planck-scale modifications of the
dispersion relation and Planck-scale modifications of translation transformations.

\subsection{Covariant mechanics in de Sitter spacetime}
\label{sec:deSitter}

In this section we present a covariant Hamiltonian formulation for the motion of a classical point particle in de Sitter spacetime.
Taking $E,p,N$ to be respectively the generators of time translations, space translations, and boosts, the algebra of spacetime symmetries for 2D de Sitter spacetime can be described in terms of Poisson brackets as
\begin{equation}
\left\{ E,p\right\} =Hp, ~~
\left\{ N , E \right\} =p+HN, ~~
\left\{ N , p\right\} =E.
\label{algebraDS}
\end{equation}
The spacetime-symmetry generators leave invariant the Casimir
\begin{equation}
{\cal C} = E^2 - p^2 -2 H N p.
\label{casimirDS}
\end{equation}

Let us then consider the ``conformal time coordinatization''
with the spatial coordinate $x$
the conformal-time coordinate $\eta$, related to the comoving time $t$ by
$$\eta = H^{-1}\left(  1 - e^{-Ht}\right) \, ,$$
 and their canonically conjugate variables $\Omega,\Pi$:
\begin{gather}
\left\lbrace \Omega, \eta \right\rbrace = 1, ~~~ \left\lbrace \Omega, x \right\rbrace = 0, ~~~ \left\lbrace \Omega, \Pi \right\rbrace = 0, \nonumber \\
\left\lbrace \Pi, \eta \right\rbrace = 0, ~~~ \left\lbrace \Pi, x \right\rbrace = -1, ~~~ \left\lbrace \eta, x \right\rbrace = 0.
\label{CanPhSpConf}
\end{gather}
We can represent the symmetry generators (\ref{algebraDS}) in terms of conformal-time coordinates as
\begin{gather}
E= \Omega (1- H \eta) + H x \Pi, ~~ p=\Pi, \nonumber \\
N = x \Omega (1-H\eta) -\Pi  \left(\eta -\frac{ H}{2} \eta^2 - \frac{H}{2} x^2\right).
\label{representDS}
\end{gather}

The invariant ${\cal H} = {\cal C} - m^2$, with $m$ the mass of the particle, can be taken to describe the particles on-shell relation (${\cal H}=0$), and then can be used as Hamiltonian constraint, in the spirit of a ``covariant formulation" of classical mechanics, generating equations of motion in terms of an auxiliary affine parameter, which we denote by $\tau$.
In terms of conformal-time coordinates (\ref{representDS}) the mass-shell relation ${\cal H}=0$ assumes the ``conformal'' aspect
\begin{equation}
{\cal H} =  (1-H\eta)^2 \left( \Omega^2 -\Pi^2\right) - m^2 = 0.
\label{on-shell-dS}
\end{equation}
The fact that the Poisson brackets between ${\cal H}$ and $E,p,N$ vanish, implies that $E,p,N$ are conserved charges, i.e. they are constant along the evolution generated by the Hamiltonian $\dot{E}=\dot{p}=\dot{N}=0$, where $\dot{f} = d f / d \tau = \{{\cal H}, f\}$. Then $E,p,N$ generate symmetry transformations (time and space translations and boosts) on the system by Poisson brackets.

One can derive the velocity of particles from the Hamiltonian constraint and the Poisson brackets~(\ref{CanPhSpConf}). Indeed by the chain rule\footnote{One can show, using the Poisson brackets~(\ref{CanPhSpConf}) and the chain rule that the velocity can be also expressed, after enforcing ${\cal H}=0$, as $v(\eta) \equiv \partial\Omega\left( \Pi \right)/\partial \Pi$.}
\begin{equation}
v(\eta) \! = \frac{dx(\eta)}{d\eta} \!= \frac{dx/d\tau}{d\eta/d\tau}\Big|_{{\cal H}=0} \!= \frac{\left\lbrace {\cal H},x\right\rbrace}{\left\lbrace {\cal H}, \eta\right\rbrace}\Big|_{{\cal H}=0} .
\end{equation}
By using Eqs.~(\ref{on-shell-dS}) and~(\ref{CanPhSpConf}) one thus finds the velocity in conformal-time coordinates (assuming $\Pi >0$)
\begin{equation}
 v(\eta) = \frac{\Pi}{\sqrt{\Pi^2 - \frac{m^2}{\left(  1 - H\eta\right)^2}}} ~~~ \stackrel{m\rightarrow 0}{\longrightarrow} ~~~ 1.
 \label{velocityConf}
\end{equation}
Notice that in the equivalent description given in terms of  the comoving time the velocity is
\begin{equation}
v(t) \! = \! \frac{dx}{dt} \!=\! e^{\!-Ht} v(\eta(t)) \!=\! \frac{ e^{-Ht} \Pi}{\sqrt{\Pi^2 \!-\! e^{2Ht} m^2 }}  \stackrel{m\rightarrow 0}{\longrightarrow} e^{-Ht} .
 \label{velocityCom}
\end{equation}

\subsection{Preliminaries on Planck-scale deformations of translations}
As announced, we shall find that there are important issues at stake particularly in the understanding of how Planck-scale effects may affect translation transformations. It is valuable for us to provide early on in this manuscript some intuition to our readers concerning the reason why these  aspects of translation transformations become so significant when spacetime curvature is  taken into account, while they are instead largely irrelevant in the Minkowski limit. We shall be satisfied in this subsection with offering only a few remarks and exclusively for the DSR-relativistic case.
In the following sections the relevant issues will of course be analyzed
in detail, both for the DSR-relativistic case and the LIV case.

In the Minkowski limit the realm of possibilities we could consider is rather tightly constrained by the fact that the characteristic scale of quantum-gravity effects  with the dimensions of an inverse-momentum (which we denote by $\ell$ for DSR), is the only scale available for deforming the mass-shell relation. This in particular implies that at leading order in this deformation scale one can only have two possible new terms, one proportional to the cubic power of energy  $\ell E^3$ and one linear in energy and quadratic in momentum  $\ell E p^2 $, so that\footnote{As usually done in this research area, we exclude terms going with odd powers of momentum, since (in their application to spacetimes with two or more spatial dimensions) they would bring in implications also for spatial rotations, implications which are not usually expected in the quantum-gravity literature.}
$$m^2=E^2-p^2+\alpha\ell E^3+\beta\ell Ep^2$$

It is then easy to see why such modifications of the mass-shell relation have no impact on translation transformations in the Minkowski limit: both the correction term  $\ell E^3$ and the correction term
 $\ell E p^2 $ have vanishing Poisson brackets with the generators of translations, which are $E$ and $p$ themselves, and have vanishing Poisson brackets among them in the Minkowski limit:
$$\{E,p\}=0$$
so that the modified mass-shell is still invariant under the standard action of space and time translation generators.

As we underlined in the previous subsection, the situation is very different when
spacetime is curved: the translation generators no longer have vanishing Poisson brackets, as shown by Eq.~(\ref{algebraDS})
$$\{E,p\}=Hp \, .$$

It turns out that in the DSR case and in presence of spacetime curvature, the assumption of undeformed translation transformations is only compatible with the term $\ell E p^2$ whereas allowing for the term
  $\ell E^3$ imposes a rather severe modification of translation transformations.

  Since the quantum-gravity literature provides no argument favouring the term  $\ell E p^2$ over  the term
  $\ell E^3$ evidently any systematic study of quantum-gravity modifications of the mass-shell relation should take into account what we here find on the implications of the translation sector.

\section{LIV with constant curvature}
\label{sec:LIVdeSitter}

Evidently the general idea of a LIV modification of de Sitter kinematics
could be realized in an infinity of ways: if one allows  breaking
relativistic symmetries, as is the premise of the LIV scenario, the framework
is totally unconstrained. It should be understood that a fully general
analysis of the LIV scenario is accordingly impossible.
It is nonetheless useful for our purposes to consider at least a few
possibilities, also as a way to show that the differences between alternative
LIV scenarios are not ``merely academic'' but rather have tangible (potentially
observable) consequences. This objective is pursued efficiently
by taking as starting point for the LIV modification the de Sitter property
coded in (\ref{on-shell-dS}) and adding a few parametrized terms
of LIV type:
 \begin{equation}
\begin{split}
 {\cal H} \!= & -m^2 + (1 \!-\! H\eta)^2 \!\left( \Omega^2 \!-\! \Pi^2\right) \!+\! \left( \tilde{\alpha} \Omega^3 \!\!+\! \tilde{\beta} \Omega \Pi^2 \right) \! \times\\ &
\!\!\!\!\!\!\!\!\!\!\!\!\!\!\! \times \!\! \left[ \! \lambda' \!\! \left( 1 \!-\! H\eta \right) \!+\! \lambda'' \!\! \left( 1 \!-\! H\eta \right)^2 \!\!+\! \lambda \! \left( 1 \!-\! H\eta \right)^3 \!\!+\!\! \lambda''' \!\! \left( 1 \!-\! H\eta \right)^4  \! \right]\!\!.
\end{split}
\label{dispLIVrep}
\end{equation}
Here $\lambda'$, $\lambda''$, $\lambda$\ and $\lambda'''$ are different choices of the inverse-momentum
scale characterizing the contribution of different LIV terms, which
 differ essentially for their time dependence, while $\tilde{\alpha},\tilde{\beta}$ are parameters governing the dependence of these terms on the particle's energy and momentum.
 Of course, one could add not only other forms of time dependence of the LIV terms, but also LIV terms of completely different form (for example involving
 spatial dependence),
 but these few terms we introduced will suffice for exposing the strength
 of our concerns.

Important for the main objective of this study is the fact that the
 LIV modification coded in (\ref{dispLIVrep}) is not
in general translationally invariant.
One can check by using (\ref{algebraDS}) and (\ref{CanPhSpConf}) that (\ref{dispLIVrep}) is invariant under the action of spatial translations $p$, but is not in general invariant under the action of time translations generated by $E$. The Poisson bracket between $E$ and ${\cal H}$ is
\begin{equation}
\begin{split}
& \left\{{\cal H}, E\right\} = H \left(\tilde{\alpha}\Omega^{3}
+\tilde{\beta}\Omega\Pi^{2}\right) \times \\ &
\times \! \left(\! -2\lambda'\left(1 \!-\! H\eta\right)
-\lambda''\left(1 \!-\! H\eta\right)^{2}
+\lambda'''\left(1 \!-\! H\eta\right)^{4}\right)
\end{split}
\label{joclambda}
\end{equation}
Thus, in general, not only (evidently) boost symmetry but also
 (more implicitly) time-translational symmetries are broken by a LIV
 scenario.
Among the LIV parameters we introduced only $\lambda$ is not present
in (\ref{joclambda}), meaning that $\lambda$ is compatible with
translational invariance, while all other parameters ($\lambda'$,$\lambda''$,$\lambda'''$) are incompatible with translational
invariance. These differences are particularly striking since from
the ``LIV perspective" (the perspective of Lorentz Invariance Violation)
the four parameters $\lambda$,$\lambda'$,$\lambda''$,$\lambda'''$ should
be viewed at exactly the same level, since their Lorentz-transformation
properties are completely analogous.

As a way to quantify the differences produced by different LIV parameters
we perform a travel time analysis.
We start by noticing that
from the Hamiltonian constraint (\ref{dispLIVrep}), for massless particles, one gets the velocity (for $\Pi>0$)
\begin{equation}
 v(\eta) \! =\!  1 \!-\! \Pi \! \left(\!\! \frac{\lambda'}{\left(1 \!-\! H\eta\right)} \!+\!\! \lambda'' \!\!\!+\! \lambda \! \left(1 \!-\! H\eta\right) \!+\!\! \lambda''' \!\!\left(1 \!-\! H\eta\right)^{2} \!\! \right)
 \label{velocityLIVdeSitter}
\end{equation}
where we have set the parameters $\tilde{\alpha} \!+\! \tilde{\beta} \!=\! 1$, noticing that for massless particles they  can be both reabsorbed in the definition of the parameters $\lambda$,$\lambda'$,$\lambda''$,$\lambda'''$.

In looking for the travel time for a particle to go from one observer to another it is important to notice that the breakdown of (time) translational invariance implies that the description of a particle's Hamiltonian given in~(\ref{dispLIVrep}), and the velocity
law~(\ref{velocityLIVdeSitter}), can only hold for one of the two
distant observers: if~(\ref{dispLIVrep}) holds for Bob then Alice's
description is given by the non-trivially  translated version of~(\ref{dispLIVrep}).
Suppose that (\ref{dispLIVrep}) and (\ref{velocityLIVdeSitter}) hold in the frame of Bob who is local to a detector, and suppose that this detector reveals two photons, one ``soft'' (whose energy is small enough that the LIV effects are negligible) and one ``hard'', emitted simultaneously at a distant source, local to an observer, say Alice, at rest relatively to Bob ({\it i.e.} Alice
and Bob are related by a pure translation).
The relation between Bob and Alice's coordinates, connected by a finite deSitter-translation, can be derived by exponentiating the action by Poisson brackets of the translation generators $E,p$, Eqs.~(\ref{representDS}) and~(\ref{CanPhSpConf}). Taking Bob to be connected to Alice by a finite spatial translation followed by a finite time translation, we write symbolically
\begin{equation}
\left( \eta,x \right)^B= e^{-\xi p}\triangleright e^{-{\zeta}E}\triangleright \left( \eta,x \right)^A,
\label{TTR}
\end{equation}
where $\zeta$ and $\xi$ are the finite translation parameters, and $\triangleright$ stands for the action by nested Poisson brackets\footnote{For a generator $G$ with parameter $a$, the finite action on a coordinate $x$ is $e^{{a}G} \triangleright x \equiv \sum_{n=0}^{\infty} \frac{a^n}{n!}\left\lbrace G,x\right\rbrace_n $, where $ \lbrace G,x \rbrace_n = \lbrace G, \lbrace G,x  \rbrace_{n-1}\rbrace$, $\lbrace G, x \rbrace_0 = x $.
In this formalism, the composed action of a spatial translation followed by a time translation is given by $e^{-\xi p}\triangleright e^{-{\zeta}E}\triangleright x$ (cfr. also~\cite{DSR-DS}).}.
One finds
\begin{equation}
\begin{gathered}
 \eta^{B} \!=\! e^{H\zeta} \eta^A - \frac{e^{H{\zeta}} \!-\! 1}{H}, \qquad
 x^{B} = e^{H{\zeta}} (x^{A} - \xi ), \\
\Omega^{B} =  e^{-H{\zeta}} \Omega^{A}, \qquad
\Pi^{B} = e^{-H{\zeta}} \Pi^{A},
\end{gathered}
\label{TransCoord-dS}
\end{equation}

On the basis of~(\ref{velocityLIVdeSitter}) one deduces that Bob describes the photons trajectories to be
\begin{equation}
x^B(\eta^B) = x^B_{O_A} + \int_{\eta^B_{O_A}}^{\eta^B} v^B(\eta) d\eta,
\label{trajectoryLIVdeSitter}
\end{equation}
where $v^B$ is given by Eq.~(\ref{velocityLIVdeSitter}) written in Bob coordinates, and $x^B_{O_A},\eta^B_{O_A}$ are the coordinates that Bob assigns to the event of emission, which coincides with Alice's origin, i.e. $\eta^B_{O_A} = \eta^B(x^A=0,\eta^A=0)$, $x^B_{O_A} = x^B(x^A=0,\eta^A=0)$.
Suppose that the event of detection of the soft photon coincides with Bob frame's origin ($x^B = \eta^B = 0$). Since for the soft photon $v(\eta)\simeq 1$, (\ref{trajectoryLIVdeSitter}) implies that $x^B_{O_A}=\eta^B_{O_A}$. One then obtains the time Bob assigns to the arrival of the hard photon by setting to zero Eq.~(\ref{trajectoryLIVdeSitter}), and solving for $\eta^B$. From~(\ref{trajectoryLIVdeSitter}) and~(\ref{velocityLIVdeSitter}), and using that $x^B_{O_A}=\eta^B_{O_A}$, one finds that the hard photon is detected with a delay
\begin{equation}
 \Delta\eta \!=\! p^B_h \!\! \left[\lambda'T \!+\! \lambda''\frac{e^{HT} \!\!\!-\! 1}{H} \!+\!\lambda\frac{e^{2HT} \!\!\!-\! 1}{2H} \!+\! \lambda'''\frac{e^{3HT} \!\!\!-\! 1}{3H}\right]
 \label{delayLIVdeSitter}
\end{equation}
where we called $p^B_h$ the hard particle's momentum measured by Bob, we used Eq.~(\ref{TransCoord-dS}) to derive $\eta^B_{O_A}$, and we called $T=\zeta$ the (comoving) time distance between Alice and Bob.
We can express the delay in terms of the redshift of the source (relative to Bob) $z \!=\! -\! H \eta^B_{O_A} \!=\! e^{HT} \!\!-\! 1$:
\begin{equation}
\!\!\!\!\!\!\!\!\!\!\!
\Delta\eta \!=\! \frac{p^B_h}{H} \!\! \left[\lambda' \! \ln\left(1 \!+\! z\right) \!+\!\! \lambda'' \! z \!+\!\! \lambda \! \left( \! z \!+\! \frac{z^{2}}{2} \!\right) \!\!+\!\! \lambda''' \!\! \left( \! z \!+\! z^{2} \!+\! \frac{z^{3}}{3}\right) \!\right]
\label{dsLIVmain}
\end{equation}
where we also expressed the delay in terms of comoving time. Notice that, since $\Delta \eta = O(\ell  \Pi)$, it follows that
\begin{equation}
\Delta t = \int_0^{\Delta \eta} d\eta a(\eta) = \Delta \eta
+ O\left(\ell ^2\right),
\label{DtDeta}
\end{equation}
so that the expression of the delay in conformal and comoving time coincide at the level accurately described by the approximations we are relying on.

The term in the delay (\ref{dsLIVmain}) proportional to $\lambda$ coincides with the one advocated by Jacob and Piran in~\cite{jacobpiran}. Jacob and Piran provided
as motivation an intuitive argument in favour of switching on only the $\lambda$ parameter in LIV phenomenology. Here we exposed the fact
that the Jacob-Piran intuition unknowingly reflected a preference for assuming that
the quantum-gravity effects leave translations unaffected\footnote{The Jacob-Piran proposal~\cite{jacobpiran}
applies to the general case of an arbitrary expansion rate, but it is for us particularly
insightful to analyze it in the special case of constant expansion rate. With constant expansion rate the original theory is fully translationally invariant, whereas for non-constant expansion rate
time translations are not a symmetry. It is noteworthy that the Jacob-Piran proposal is such that
the LIV effects are fully translationally invariant (including invariance under time translations)
when the expansion rate is constant, whereas other possible LIV terms would not have this property. This
is a key aspect of what we label as a case where translations are unaffected by the quantum-gravity
effects.}.
However, as stressed above, the quantum-gravity literature providing motivation for LIV research
does not justify the assumption that quantum-gravity effects should leave translations unaffected.
This makes us particularly concerned for the fact that limits claiming
general applicability to LIV scenarios have been presented
in the literature \cite{Bolmont:2010np,Ellis:2011ek,HESS:2011aa,Aharonian:2008kz,Albert:2007qk,Abdo:2009zza,Ackermann:2009aa}, even though they were based exclusively on the Jacob-Piran {\it ansatz}. The observations reported in this section, strengthened by what we shall find in the following sections, show instead
that those limits only apply to a particular case of LIV. They should be viewed as ``conditional limits", the condition being indeed that translations are unaffected by the quantum-gravity effects producing the breakdown of Lorentz invariance.

In closing this section we offer one more remark that can provide some intuition for the significance
of our concerns, a remark which applies to the description of the equations of motion. For this we consider  a third observer Bob$'$, at rest relatively to Alice and Bob, whose origin is along the soft photon worldline connecting Alice and Bob, at some point between Alice and Bob, a point which is at (comoving) time $T'$ from
the origin of the reference frame of Bob.
 The coordinates of Bob$'$ will be related to the ones of Bob by expression~(\ref{TransCoord-dS}) replacing the coordinates of Alice with the ones of Bob$'$  and replacing $\zeta$ with $T'$. Then, from~(\ref{TransCoord-dS}) and (\ref{velocityLIVdeSitter}), it follows that Bob$'$ describes the photons traveling with velocity
\begin{equation}
\begin{split}
& v^{B'}\left(\eta^{B'}\right)=1-\Pi^{B'} \bigg(\frac{\lambda'e^{-2HT'}}{\left(1-H\eta^{B'}\right)}+\lambda''e^{-HT'} \\ &
~~~~~~~~~ + \lambda\left(1-H\eta^{B'}\right)+\lambda'''e^{HT'}\left(1-H\eta^{B'}\right)^{2}\bigg)
\end{split}
 \label{velocityLIVdeSitter-2}
\end{equation}
Here it is particularly important to notice that the only term which maintains the same form of~(\ref{velocityLIVdeSitter}) is the one proportional to $\lambda$, which indeed corresponds to the translational invariant term in the Hamiltonian~(\ref{dispLIVrep}).

\section{DSR-relativistic picture and relative locality with constant curvature}
\label{sec.DSR-dS}

In the DSR approach, the (inverse-momentum) scale $\ell$ at which the dispersion relation is modified is a relativistic invariant (it plays a role completely analogous to the role of $c$, the velocity of light, in ordinary special relativity). This requirement enforces the ($\ell$-deformed) dispersion relation to be expressible as a combination of only the (charges) generators of the relativistic symmetries; in fact, it must be a Casimir of an algebra of charges/generators. As first observed in Ref.~\cite{gacdsr1IJMPD} (also see Refs.~\cite{jurekDSRfirst,joaoleePRDdsr}), the relativistic invariance of
 a modified dispersion relation requires that the algebra of relativistic-symmetry generators must also be deformed by the scale $\ell$.

In light of these considerations, it is preferable, for the purposes of the DSR analysis, to start from a modification, rather then of the dispersion relation~(\ref{on-shell-dS}), of the Casimir equation (\ref{casimirDS}). The motion of particles is then generated by the deformed Hamiltonian constraint ${\cal H} = {\cal C} - m^2$.
We take the following deformation\footnote{Consistently with our objectives, we enforce the requirement that in
the ``flat limit'', $H\rightarrow 0$, one should get
the velocity $v \simeq 1 - \ell E$.
We also do not make room for deformation terms proportional to $N$, since our
flat-limit requirements would only allow such terms to be of the
type $\ell H N E p$ or $\ell H N E^2$, producing negligibly small effects.
In general we should proceed with full awareness of the smallness of
 both $\ell$ and $H$, so among possible terms going with $\ell H$
 we shall only keep track of those involving $\ell H x $ or $\ell H t$,
 {\it i.e.} cases where the cosmological travel times and travel distances
 here of interest can compensate for the smallness of $\ell H$.}
\begin{equation}
{\cal C}=E^{2}-p^{2}-2H N p+\ell\left(\alpha E^{3}+\beta Ep^{2}\right) ~,
\label{casimirDSR-DS}
\end{equation}
The following $\ell$-deformed (2D) de Sitter algebra of charges is compatible with the invariance of ${\cal C}$:
\begin{gather}
\left\{ E,p\right\} =Hp-\ell (\alpha -\gamma) HEp \ ,\nonumber \\
\left\{ N , E \right\} =p+HN-\ell (\alpha -\gamma) E(p+HN)-\ell\beta Ep\ ,\nonumber \\
\left\{ N , p\right\} =E+\frac{1}{2}\ell (\alpha + 2 \gamma) E^{2}+\frac{1}{2}\ell\beta p^{2}\ .
\label{algebraDSR-DS}
\end{gather}
This generalizes the results of Ref.~\cite{DSR-DS}, and reproduces
the results of Ref.~\cite{DSR-DS} for $\gamma = 0$.

Notice that, as stressed already earlier in this manuscript, the deformation term going like $E p^2$
does not require a modification of translation transformations while for the  deformation term going like $E^3$  a modification of translation transformations is required. This is particularly clearly by looking at Eqs.(\ref{casimirDSR-DS})-(\ref{algebraDSR-DS}): the parameter $\alpha$ of the term going
like $E^3$ has implications not only for transformations involving boosts ($\alpha$ affects $\left\{ N , E \right\}$ and $\left\{ N , p\right\}$) but also for
transformations involving exclusively translations ($\alpha$ also affects $\left\{ E,p\right\}$), whereas the parameter $\beta$ of the term going like $E p^2$ does not affect $\left\{ E,p\right\}$.

To study the motion of particles, we want to express the deformed
Hamiltonian ${\cal H} = {\cal C} - m^2$ in terms of a set of spacetime coordinates and conjugate momenta, and derive the particles velocity in terms of these coordinates, as done in Subsec.~\ref{sec:deSitter} and Sec.\ref{sec:LIVdeSitter}. At this stage one must expect a crucial difference with respect to the LIV case,
 a feature known in the literature as ``relative locality"~\cite{principle} (also see Refs.\cite{bob,bobKAPPAplb}.
  This is another consequence of the fact that, in the DSR case, the algebra of the generators of spacetime relativistic-symmetry transformations is deformed by the inverse-momentum scale $\ell$: the result is that the action of the whole set of the symmetry generators on spacetime coordinates must be necessarily momentum-dependent, leading to the presence of
  momentum-dependent misleading inferences for the description
  by a given observer of events occurring at a {\underline{large distance}} from that observer (large distance from the origin of the reference frame of that observer). This core feature of ``relative locality" ~\cite{principle} is usually
  rather surprising for newcomers, but actually it can be easily understood
  via an analogy with the one case of deformation of relativistic symmetries
  we all know, the one of the deformation from Galileian to special-relativistic symmetries: according to the mindset of a Galileian-era
  physicist the special-relativistic transformations introduce misleading
  inferences concerning the simultaneity of events, misleading inferences
  of increasing severity when the compared notions of simultaneity concern
  observers connected by a {\underline{large boost}}.

  We shall deal with these challenges drawing from previous results on relative locality, but we shall at the same time provide sufficient details for a self-contained description of the relative-locality effects.

 We start by seeking a representation\footnote{Previous related studies of relative locality have shown~\cite{bobKAPPAplb,anatomy,DSR-DS} that the physical results do not depend on the choice of the representation.} of the charges~(\ref{algebraDSR-DS}) suitable for a comparison with the LIV scenario of Sec.~\ref{sec:LIVdeSitter}. We choose the following representation in terms of the canonically conjugate  ``conformal-time coordinates" $\Omega,\Pi,\eta,x$ defined in (\ref{CanPhSpConf}):
\begin{equation}
\begin{split}
& \!\! E \! = \!(1 \!-\! H \eta) \Omega \!+\! H x \Pi \!-\!\! \frac{\ell}{2} \!\left( \alpha \!-\! \gamma \right) \! \left( (1 \!-\! H \eta) \Omega \!+\! H x \Pi \right) \!,
\\ & p=\Pi, \\ &
N = x \Omega (1-H\eta) -\Pi  \left(\eta -\frac{ H}{2} \eta^2 - \frac{H}{2} x^2\right) \\ &
+\frac{\ell}{2} \beta \left(\eta \left(2 - 3 H\eta + H^2\eta^2 \right) \Omega \Pi  + x \Pi ^2\right) \\ &
\!+\! \frac{\ell}{2} \gamma  x \left(H^2 x^2 \Pi^2 \!\!+\! 3 \Omega (1-H\eta) \left(\Omega (1 \!-\! H\eta) \!+\! H x \Pi \right)\right)
\end{split}
\label{repDSR-DS}
\end{equation}
With this representation, the Casimir~(\ref{casimirDSR-DS}) takes the form
\begin{equation}
{\cal C} \!=\! (1-H\eta)^2 \!\left( \Omega^2 \!-\! \Pi^2\right) \!+ \ell \left( 1 \!-\! H\eta \right)^3 \! \left( \gamma \Omega^3 \!\!+\! \beta \Omega \Pi^2 \right)\!,
\end{equation}
and it is interesting to notice that the corresponding form of the Hamiltonian constraint ${\cal H} = {\cal C} - m^2$ reproduces the one in the LIV scenario~(\ref{dispLIVrep}) for  the special choice $\lambda \neq 0$, $\lambda' = \lambda'' = \lambda'''=0 $,  which is the particular LIV case in which translational symmetry is not broken.  Much insight  on the differences between the LIV\ and the DSR\ scenarios will be here provided by contrasting these two cases with the same Hamiltonian constraint. We shall see that \\
$\star$ even when they rely on the same Hamiltonian constraint a LIV and a DSR scenario do not in general lead to the same predictions,\\
$\star$ and in particular even a case where the Hamiltonian constrain requires nothing new of translation transformations in the LIV scenario can require in the DSR scenario a deformation of translation-symmetry transformations.

This last point is already evident in (\ref{repDSR-DS}) which shows how the representation of (time-)translation generators depends in the DSR case on the parameters $\alpha$ and $\gamma$, through the combination $\alpha - \gamma$.
So translations are unaffected by the parameter $\beta$ but do depend on $\gamma$, the other parameter present in the Hamiltonian constraint. They also depend on $\alpha$ which is a parameter coding properties specifically of boost-symmetry transformations.

For massless particles, setting to zero the Hamiltonian constraint~(\ref{casimirDSR-DS}), one finds the velocity (for $\Pi>0$)
\begin{equation}
v(\eta) = 1 - \ell \left( \gamma + \beta \right)\left( 1- H\eta \right) \Pi
\label{velocityDSRdS}
\end{equation}
Differently from the LIV scenario of the previous section, in the DSR scenario the (deformed) relativistic symmetries are preserved, and all the relativistic observers describe (massless) particles moving, in their coordinates, with the same expression for the velocity, given in Eq.~(\ref{velocityDSRdS}). Indeed the generators of relativistic-symmetry transformations,  $E,p$ and $N$, are conserved charges, and therefore their Poisson bracket with the Casimir/Hamiltonian is null, as one can easily verify.

We are interested, as in Sec.~\ref{sec:LIVdeSitter}, in the time of arrival at a distant detector of two photons, one ``hard" ($x_h,\eta_h,p_h$) and one ``soft" ($x_s,\eta_s,p_s$), emitted simultaneously at a distant source. Again, take Alice and Bob to be the observers local respectively to the source and to the detector.
As in~(\ref{trajectoryLIVdeSitter}), Bob describes photons to move along the trajectories
\begin{equation}
\begin{split}
 x^B(\eta^B) & =  x^B_{O_A} \!+\!\! \int_{\eta^B_{O_A}}^{\eta^B} \!\!\!\!\! d\eta\ v^B(\eta) = x^B_{O_A} \!+\! (\eta^B \!\!-\! \eta^B_{O_A}) \\ &
~ -\! \ell (\eta^B \!\!-\! \eta^B_{O_A}) (\gamma \!+\! \beta)\Pi^B \!\! \left( 1 \!-\! H(\eta^B \!\!+\! \eta^B_{O_A}) \right)
\end{split}
\label{worDSRdS}
\end{equation}
where $v^B$ now is given by Eq.~(\ref{velocityDSRdS}).
For simplicity we focus on the case such  that the soft photon reaches the detector in the origin of Bob's frame, i.e. its worldline crosses $\eta^B_s = x^B_s =0$.
In order to derive the measured time of arrival at the detector of the hard photon we need to enforce the event of emission to be local for Alice, who is at the source. This is most clearly investigated by focusing on the case such that  both particles are emitted, according to Alice, at $\eta^A = x^A = 0$. As mentioned above, differently from the LIV case, since the translations~(\ref{PhSpDSRdS}) depend on momenta, they are affected by relative locality. Then the coordinates $x^B_{O_A} \!\!\!=\! x^B \!(x^A\!\!\!= \!\!0,\eta^A \!\!\!= \!\!0)$, $\eta^B_{O_A} \!\!\!=\! \eta^B \!(x^A\!\!\!= \!\!0,\eta^A \!\!\!= \!\!0)$ that Bob assigns to the (distant) events of emission do not coincide, as a result of the fact that the  particles have different momenta (and, as stressed and shown above, translation transformations act in momentum-dependent manner). In order to evaluate $\eta^B_{O_A}, x^B_{O_A}$ we need to calculate the finite translations connecting Alice and Bob. As shown in~\cite{DSR-DS}, the relation between Bob and Alice's coordinates can be derived by exponentiating the action by Poisson
brackets of the translation generators $E,p$, which is
\begin{equation}
\begin{gathered}
\left\lbrace E, \eta \right\rbrace = (1-H\eta) \left( 1 - \ell (\alpha -\gamma) E \right), ~~~ \left\lbrace p, \eta \right\rbrace = 0, \\
\left\lbrace E, x \right\rbrace = - H x \left( 1 - \ell (\alpha -\gamma) E \right), ~~~ \left\lbrace p, x \right\rbrace = -1.
\end{gathered}
\label{PhSpDSRdS}
\end{equation}

 As in Sec.~\ref{sec:LIVdeSitter},
taking Bob to be connected to Alice by a finite spatial translation followed by a finite time translation, Eq.~(\ref{TTR}), one finds
\begin{gather}
 \eta^{B} \!=\! \frac{1 \!-\! e^{H{\zeta}}}{H} \!+ e^{H{\zeta}}\eta^{A}
 \!+ \ell\,(\alpha \!-\! \gamma) \, {\zeta} e^{H{\zeta}} (1 \!-\! H\eta^{A}) E^{A}_{H,\xi}, \nonumber\\
 x^{B} = e^{H{\zeta}} (x^{A} - \xi) - \ell (\alpha \!-\! \gamma) {\zeta} e^{H{\zeta}} H(x^{A}-\xi) E^{A}_{H,\xi},
\nonumber\\
\Omega^{B} =  e^{-H{\zeta}} \left(\Omega^{A} + \ell (\alpha \!-\! \gamma) H {\zeta}  \Omega^{A} E^{A}_{H,\xi}\right), \nonumber\\
\Pi^{B} = e^{-H{\zeta}}\left(\Pi^{A} + \ell(\alpha \!-\! \gamma) H{\zeta} \Pi^{A}E^{A}_{H,\xi} \right),
\label{TransCoord-DSR}
\end{gather}
where we found convenient to introduce
\[E^{A}_{H,\xi} \equiv \Omega^A\left( 1-H\eta^A \right)+H\Pi^A\left( x^A -\xi\right)\]

Notice that the condition for Bob's origin to coincide with the event of detection of the soft photon is enforced by setting $\zeta = T$ and $\xi \!=\! H^{-1}(1 \!-\! e^{-HT})$, where $T$ is the (comoving) time distance between Alice and Bob.
From these relations one gets (also using the fact that, for a massless particle on shell, one has $\Omega = \Pi + O(\ell \Pi)$)
\begin{equation}
\begin{gathered}
\eta^{B}_{O_A} \!=\! - \frac{e^{H{T}} \!-\! 1}{H} \!+ \ell\,(\alpha \!-\! \gamma) \, {T} e^{H{T}} \Pi^B \\
x^{B}_{O_A} =  - \frac{e^{HT} \!-\! 1}{H} + \ell (\alpha \!-\! \gamma) {T} \left(  e^{HT} \!-\! 1\right) \Pi^B.
\end{gathered}
\end{equation}
Substituting $\eta^B_{O_A}, x^B_{O_A}$ in~(\ref{worDSRdS}) and solving for $\eta_h^B(x^B_h \!\!=\! 0)$ we get the time of arrival at the detector of the hard photon:
\begin{equation*}
\!\!\!\Delta \eta^B \!\!=\! \eta_h^B(x^B_h \!\!=\! 0) \!=\! \ell p_h^{B} \!\! \left( \!\!(\alpha \!-\! \gamma) T \!\!+\! (\beta\!+\!\gamma)\frac{e^{2H T}\!\!\!-\!1}{2H} \!\right).
\label{delayBOBold}
\end{equation*}
At this stage it is convenient to remove the suffixes indicating the observer and express the detected delay in terms of the redshift $z$ characterizing the source, giving us the result in the form
\begin{equation}
\Delta t=\ell p_h \left(\left( \alpha - \gamma \right)\frac{\ln\left(1+z\right)}{H}+(\beta+\gamma)\frac{z+\frac{z^{2}}{2}}{H}\right),
\label{delayBOBz}
\end{equation}
where we denoted by $p_h$ the momentum of the hard particle observed at the detector, and we expressed the delay in terms of comoving time (keeping again in mind the observation reported in (\ref{DtDeta})).

\section{FRW spacetime combining slices of de Sitter spacetime}
\label{sec.Slicing}

So far we focused on spacetimes with constant curvature (de Sitter), providing for the LIV case a generalization of the results of Ref.~\cite{jacobpiran} and providing for the DSR case a genealization of the results of Ref.~\cite{DSR-DS}.
For our purposes it is crucial to further generalize these result to the case of non-constant curvature, specifically the case of FRW-type expansion, since our analysis is aimed at applications of  modified dispersion relations in astrophysics and cosmology. For the constant-curvature case our strategy of analysis proved to be very powerful  thanks to its reliance on techniques that exploit the high amount of symmetries  present in the de Sitter case. This is of course particularly evident for DSR-relativistic scenarios, but also plays a role for the LIV\ scenarios (where the relativistic symmetries are broken, but at zero-th order of analysis one has them all). The fact that for FRW expansion one looses quite a bit of those symmetries poses a challenge from our perspective. We shall argue however that the connection between de Sitter spacetimes and FRW spacetimes is nonetheless strong enough to provide a clear path for the generalization we are seeking. Specifically we shall use the fact that  FRW spacetimes can be described in terms of  a suitable sequence of ``thick slices" of de Sitter spacetimes.

In the following sections we shall use this notion of ``de-Sitter slicing of FRW spacetimes" for the purpose of deriving predictions for the implications of modified dispersion relations in presence of FRW-type spacetime expasion at non-constant rate. In preparation for that we first show in this section how the ``thick slicing" can be used in the standard case of the propagation of a classical particle in a classical FRW spacetime (no modifications of the dispersion relation). In doing so we shall build on the strength of related results on ``de Sitter slicing" already reported in Ref.~\cite{interplay}. We
refine the proposal put forward
  in~\cite{interplay} by strengthening the role of symmetry generators in characterizing each de-Sitter slice and by making reference to observers associated to each slice. This refinements prove valuable in order to deal with the effects of relative locality, which require physical observations to be described by observers local to the spacetime events~\cite{principle}.

We find
that it is convenient to perform the analysis
 first in terms of the comoving time $t$.
The FRW spacetime is described by the metric
\begin{equation}
 ds^2 = dt^2 - a^2(t) dx^2,
\label{metricFRW}
\end{equation}
where $a(t)$ is the scale factor of the universe, defined by the relation $H(t) \!=\! \dot{a}(t)/a(t)$, $H(t)$ being the (time-dependent) expansion rate, from which we also get
\begin{equation}
\frac{a\left(t_{f}\right)}{a\left(t_{i}\right)}=\exp\left( \int_{t_{i}}^{t_{f}}dt\ H\left(t\right) \right)
\label{scaleH}
\end{equation}
We reconstruct the trajectory of a FRW massless particle from its source to a distant detector considering a first observer, Alice, local to the emission, and a final observer, Bob, local to the detection.
If the particle moves in a FRW spacetime, it is straightforward from~(\ref{metricFRW}) to derive Bob's description of its trajectory as
\begin{equation}
x^B\left( t^B \right) = x^B_{O_A} + \int_{t^B_{O_A}}^{t^B} \frac{dt}{a\left( t \right)}
\label{FRWworld}
\end{equation}
where we are assuming $a(t^B=0) = 1$.

Since in this section we just consider a classical particle in a classical FRW spacetime, the ``thick slices" here of interest will have the geometry of a classical de Sitter spacetime,
which in particular affords us also the luxury of analyzing each slice using classical translational invariance (this will change in the later section for the cases where the deformation affects translational invariance).

Our ``thick slices" of FRW
spacetime are introduced by dividing the time interval between the event of emission and the event of detection in $N$ time intervals of temporal size\footnote{Above we used the symbol $\Delta t$ to indicate the delay in time of arrival of a hard photon with respect to a soft photon. Here with the symbol $\Delta t$ we indicate the thickness, in time, of each slice. Which notion of $\Delta t$
we refer to at a given point in the manuscript should always be clear from the context.} $\Delta t_n$, $n \!=\! 1,..,N$. In each slice spacetime is described,to good approximation, by a  constant expansion rate $H_n \!=\! H\left( t_n \right)$, where $t_n$ is the initial time of the $n$-th slice. For simplicity we  divide the time interval in slices of equal size $\Delta t_n \!=\! T/N$, where $T$ is the time of flight of the photon from the source to the detector. Then of course we have that $t_n \!=\! t_i + n T/N $.
Starting from the observer Alice, who is at the source, we contemplate a set of intermediate observers Bob$_n$, $n=1,..,N$, such that each $n$-th observer crosses the photon's trajectory at the time $t_n$. Each observer Bob$_n$, in the corresponding $n$-th slice, which goes from $t_{n-1}$ to $t_{n}$, will describe the motion of particles in terms of a constant expansion rate $H_n$.

We are of course ultimately interested in taking the limit $N \rightarrow \infty$, so that we have an infinity of infinitesimally small slices and the assumption of constant expansion rate in each slice becomes fully justified.

We can also describe each Bob$_n$ as the observer connected to Alice by a set of $n$ spatial translations followed by a set of $n$  time translations, with each $k$-th translation characterized by the relative constant expansion rate $H_k$ and finite translation parameters $\zeta_k,\xi_k$, i.e.
\begin{equation}
\left(t,x\right)^{B_{n}}\!\! = e^{-\sum_{k=1}^n \xi_k p}
\triangleright e^{-\sum_{k=1}^n \zeta_kE_{H_k}}\triangleright\left(t,x\right)^A \!\!.
\label{FiniteTransSlicing}
\end{equation}
These give the relation between Bob$_n$ and Alice's coordinates
\begin{equation}
\begin{gathered}
 t^{B_{n}}\left(t^{A},x^{A}\right)=t^{A}-\sum_{k=1}^n\zeta_k,\\
x^{B_{n}}\left(t^{A},x^{A}\right)=e^{\sum_{k=1}^{n}H_{k}\zeta_k}\left(x^{A}-\sum_{k=1}^{n}\xi_k\right).
\label{BobnAl}
\end{gathered}
\end{equation}
The requirement for each observer Bob$_n$ to be along the photons trajectories at the time $t_n$, is then ensured by imposing that the translation parameters satisfy the conditions
\begin{equation}
 \zeta_n = \zeta = T/N , \qquad
 \xi_n = e^{-\sum_{k=1}^{n}H_{k} \zeta_n} \frac{e^{H_{n} \zeta_n} - 1}{H_{n}},
 \label{param}
\end{equation}
and Alice describes the slices to be of sizes $\Delta t_n^A \!=\! \zeta$, $\Delta x_n^A \!=\! \xi_n$.

We are interested in the worldline described by the observer Bob$_N$, which is our Bob, local to the detection.
From the relations (\ref{BobnAl}) we can derive the relation between the coordinates of  Bob$_n$ and the coordinates of Bob$_N$
\begin{equation}
\begin{gathered}
t^{B_{N}} = t^{B_{n}}-\left(N-n\right)\zeta \\
x^{B_{N}} \!\! =\! e^{\sum_{k=n+1}^{N} \! H_{k}\zeta}x^{B_{n}} \!-\! \!\!\! \sum_{k=n+1}^{N}e^{\sum_{s=k}^{N} \! H_{s}\zeta}\frac{1 \!-\! e^{-H_{k}\zeta}}{H_{k}}.
\end{gathered}
\label{BobnBobN}
\end{equation}
Since, as seen using these relations,
\begin{equation}
v^{B_{N}} \!=\! \frac{dx^{B_{N}}}{dt^{B_{N}}} \!=\! e^{\sum_{k=n+1}^{N} \!\! H_{k}\zeta}\frac{dx^{B_{n}}}{dt^{B_{n}}} \!=\! e^{\sum_{k=n+1}^{N} \!\! H_{k}\zeta}v^{B_{n}} \!,
\end{equation}
Bob$_N$, in each $n$-th slice, describes photons to move with velocity
\begin{equation}
\begin{split}
v_{n}^{B_{N}} \!\! \left(t^{B_{N}} \! \right) & \!\!=\! e^{\sum_{k=n \!+\! 1}^{N} \!\! H_{k}\zeta} v^{B_{n}}_n \!\! \left(t^{B_{n}} \!\! \left(t^{B_{N}} \!\right) \! \right) \\
& \!\!=\! e^{\sum_{k=n \!+\! 1}^{N} \!\! H_{k}\zeta}  e^{ -H_{n} \! \left( \! t^{B_{\!N}} \! \!-t_{O_n}^{B_{\!N}} \!\right)},
\end{split}
\end{equation}
where, here and in the following, we denote by $t^{B_N}_{O_n}$ and $x^{B_N}_{O_n}$, the value of the coordinates that the observer Bob$_N$ attributes to the position of the particle at crossing  of the spatial origin of observer Bob$_n$'s frame.

Then, the photon trajectories in the $n$-th slice, are described
by Bob$_N$ to be
\begin{equation}
x^{B_{N}} \! \left(t^{B_{N}} \!\right)_{n} \!\!=\! x_{O_A}^{B_{N}} \!+\! \sum_{k=1}^{n-1} \int_{t_{O_{k-1}}^{B_{N}}}^{t_{O_{k}}^{B_{N}}} \!\!\!\! dt\ \!v_{k}^{B_{N}} \!+\!\! \int_{t_{O_{n-1}}^{B_{N}}}^{t^{B_{N}}} \!\!\!\!\!\!\! dt\ \! v_{n}^{B_{N}} .
\label{trajectoryFRW-N}
\end{equation}
Since the observers Bob$_n$ are defined so that the photons cross the origin of their reference frame, the second term on the right-hand side can be obtained using relations (\ref{BobnBobN}) with $t^{B_N}_{O_n} \!\!\!=\! t^{B_N}(t^{B_n} \!\!=\! 0) $, $x^{B_N}_{O_n} \!\!\!=\! x^{B_N}(x^{B_n} \!\!=\! 0) $, is
\begin{equation}
\sum_{k=1}^{n-1} \! e^{\sum_{s =k + 1}^{N} \!\! H_{s}\zeta} \! \frac{e^{H_{k}\zeta} \!\!-\! 1}{H_{k}} .
\label{term-1}
\end{equation}
Considering that $\zeta \!=\! T/N \!=\! \Delta T_n$, in the limit $N \! \rightarrow\infty$,
\begin{equation}
\sum_{k=n_i+1}^{n_f} \zeta \rightarrow  \int_{t_{n_i}}^{t_{n_f}} dt.
\label{Riemann}
\end{equation}
It follows from~(\ref{scaleH}) that
\begin{equation}
e^{\sum_{s=k+1}^{n}H_{s}} \rightarrow \frac{a\left( t_n \right)}{a\left( t_k \right)},
\label{Riemann-2}
\end{equation}
Then the term~(\ref{term-1}) tends to
\begin{equation}
\sum_{k=1}^{n-1}\frac{\zeta}{a\left(t_{k}\right)} \rightarrow \int_{t_{O_A}^{B_{N}}}^{t_{O_{n-1}}^{B_{N}}} \!\!\!\! \frac{dt}{a\left(t\right)}
\label{term-1-2}
\end{equation}
where we considered that for $N \!\rightarrow \! \infty $ one has that $\frac{e^{H_{k}\zeta} - 1}{H_{k}} \!\rightarrow \zeta$ and that $a\left( t_N \right) \!=\! 1$ since $t_N \!=\! t_{O_N}$.
For what concerns the last term in~(\ref{trajectoryFRW-N}), consider that in the slice $t_{O_{n-1}}^{B_{n}} \!\! \leq  t \leq t_{O_n}^{B_{n}}$,
where $H\left(t\right)=H_{n}=\text{const.}$,
\[
e^{-H_{n}\left(t-t_{O_n}^{B_{N}}\right)}\simeq\frac{a\left(t_{n}\right)}{a\left(t\right)},
\]
so that, using also~(\ref{Riemann-2}), the term becomes
\begin{equation}
e^{\sum_{s=n + 1}^{N} \!\! H_{s}\zeta} \!\! \int_{t_{O_{n-1}}^{B_{N}}}^{t^{B_{N}}} \!\!\!\!\!\!\!\!\! dt\ \!e^{- \! H_{n} \!\left(\! t - t_{O_n}^{B_{N}} \!\right)}  \rightarrow \int_{t_{O_{n-1}}^{B_{N}}}^{t^{B_{N}}} \!\!\! \frac{dt}{a\left(t\right)}
\label{term-2}
\end{equation}
and, substituting (\ref{term-1-2}) and (\ref{term-2}) in~(\ref{trajectoryFRW-N}), we obtain that as $N \! \rightarrow \infty$ the trajectory of Bob$_N$ tends to the trajectory of Bob, given in~(\ref{FRWworld}).

The fact that in the $N \! \rightarrow \infty$ limit our thick-slicing procedure would match exactly the results obtained in a FRW spacetime was assured by construction. It was useful for what follows to see, as done above, the technical details of how this convergence takes shape. On the quantitative side it is rather impressive how quickly this convergence takes place:\ as shown in Fig.1 even for relatively small values of N our slicing procedure gives results that are already in pretty good agreement with the results of the analysis done in the corresponding FRW spacetime.

\begin{figure}[h!]
\includegraphics[scale=0.9]{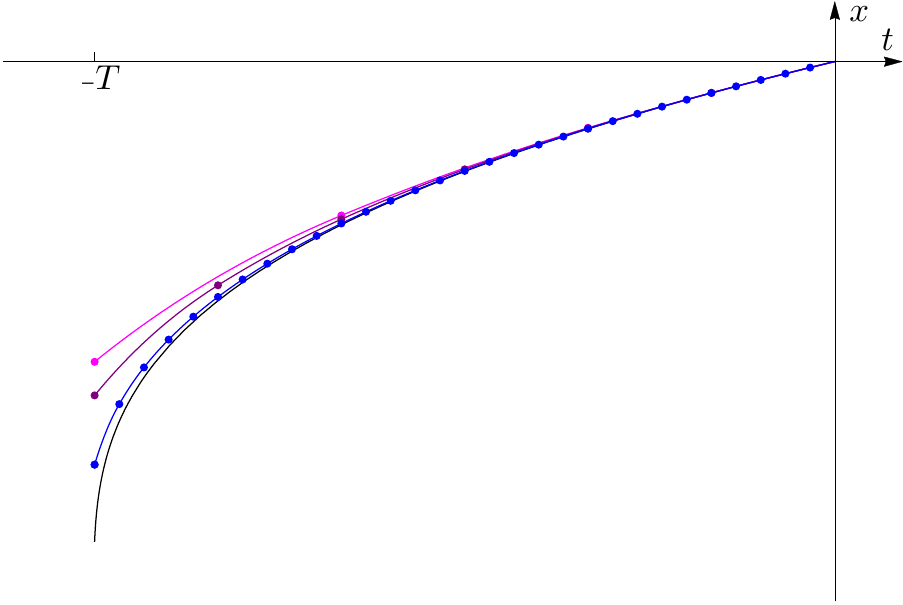}
\caption{Here we show the worldline of a massless particle emitted at a distant source at the time $-T$, from the point of view of an observer at the detector. We choose for our pictorial example the scale factor to obey the power law behavior $a(t^B) = ((t^B+t_0)/t_0)^{2/3}$, were $t_0$ is a constant indicating the ``Big-Bang'' time. We show respectively the trajectory (black curve) described by a FRW observer, and the trajectories described by observers (Bob$_N$) obtained through our thick-slicing procedure, Eq.~(\ref{trajectoryFRW-N}), with different levels of approximations (curves magenta, purple and blue). The dots indicate the spacetime position of each Bob$_n$ observer, as described by Bob$_N$. We see that as the  slicing approximation gets finer (as $N$ increases) the particle's worldline, as well as the source position, converge to the ones of FRW spacetime.}
\end{figure}

Further insight can be gained by working with conformal-time coordinates.
Conformal time is defined by the relation $dt = a(\eta)d\eta$, where $a(\eta) = a(t(\eta))$. In conformal-time coordinates, the FRW spacetime metric becomes
\begin{equation}
 ds^2 = a^2(\eta)\left(d\eta^2 - dx^2  \right).
\end{equation}
Notice also that the expansion rate can then be expressed as follows
\begin{equation}
H(t) = \frac{1}{a(t)}\frac{da(t)}{dt} = \frac{1}{a^2(\eta)}\frac{da(\eta)}{d\eta}
\label{expRate}
\end{equation}
Bob's description of the worldline~(Eq.~(\ref{FRWworld})) is
\begin{equation}
x^B(\eta^B) = x^B_{O_A} - \eta^B_{O_A} + \eta^B = \eta^B,
\label{FRWworld-CON}
\end{equation}
where the requirement for Bob's origin to coincide with the detection of the photon enforces
\begin{equation}
\eta^B_{O_A} = x^B_{O_A} = - \int_{-T}^{0} \frac{dt}{a(t)}.
\label{emissFRW}
\end{equation}

The slices are of size $\Delta\eta_n \!=\! H_n^{-1} \!\left(  1 \!-\! e^{-H_n T/N}\right)$and  the relations between Alice and Bob$_n$ coordinates, generated by the action~(\ref{FiniteTransSlicing}), are
\begin{equation}
\begin{gathered}
\eta^{B_{n}} \! =\! e^{\sum_{k=1}^n \!\! H_k \zeta_k} \eta^A \!\!-\! \sum_{k=1}^n e^{\sum_{s=k+1}^n \!\! H_s \zeta_s} \frac{ e^{H_k \zeta_k} \!-\! 1}{H_k}, \\
x^{B_{n}} \!=e^{\sum_{k=1}^n H_k \zeta_k} \left( x^A - \sum_{k=1}^N \xi_k \right).
\label{BobnAl-CON}
\end{gathered}
\end{equation}
Imposing the conditions~(\ref{param}), one finds from~(\ref{BobnAl-CON}) the relations between Bob$_n$ and Bob$_N$ coordinates:\begin{equation}
\begin{gathered}
\eta^{B_{N}} \!\!=\! e^{\sum_{k=n+1}^{N} \!\! H_{k} \zeta} \eta^{B_{n}} \!\!-\!\!\!\! \sum_{k=n+1}^{N} \!\!\! e^{\sum_{s=k+1}^{N} \!\! H_{s}a_{t}} \frac{e^{H_{k}\zeta} \!\!-\! 1}{H_{k}},\\
x^{B_{N}} \!\!=\! e^{\sum_{k=n+1}^{N} \!\! H_{k} \zeta} x^{B_{n}} \!\!-\!\!\!\! \sum_{k=n+1}^{N} \!\!\! e^{\sum_{s=k+1}^{N} \!\! H_{s}a_{t}} \frac{e^{H_{k}\zeta} \!\!-\! 1}{H_{k}} \\
\Pi_{B_{\!N}} \!\!\!=\! e^{-\!\sum_{k=n+1}^{N} \!\! H_{k}\zeta} \Pi_{\!B_{n}}, ~~~ \Omega_{B_{\!N}} \!\!\!=\! e^{-\!\sum_{k=n+1}^{N} \!\! H_{k}\zeta} \Omega_{B_{n}}
\label{BobnBobN-CON}
\end{gathered}
\end{equation}
Each Bob$_n$ describes, in each
$n$-th slice, the photons to move with velocity $v_{n}^{B_{n}} \! \left(\eta^{B_{n}}\!\right) \!=\! 1$, since, using Eq. (\ref{BobnBobN-CON}),
\begin{equation}
v^{B_{N}}=\frac{dx^{B_{N}}}{d\eta^{B_{N}}}=\frac{dx^{B_{n}}}{d\eta^{B_{n}}}=v^{B_{n}}=1,
\label{velBNBn}
\end{equation}
Then, it is straightforward to see that Bob$_N$'s description of the photons trajectories in the $n$-th slice coincides with Eq.~(\ref{FRWworld-CON}):
\begin{equation}
\begin{split}
& x^{B_{N}}\!\! \left(\eta^{B_{N}} \!\right)_{\!N} \!\!=\! x_{O_A}^{B_{N}} \!+\!\! \sum_{k=1}^{n-1} \! \int_{\eta_{O_{k-1}}^{B_{N}}}^{\eta_{O_{k}}^{B_{N}}} \!\!\!\!\! d\eta\ \! v_{k}^{B_{N}} \!\! +\!\! \int_{\eta_{O_{n-1}}^{B_{N}}}^{\eta^{B_{N}}} \!\!\!\!\!\!\!\!\!\! d\eta\ \! v_{n}^{B_{N}}\\
& = x_{O_A}^{B_{N}} \!\!+\! \sum_{k=1}^{n-1}\left(\eta_{O_{k}}^{B_{N}} \!\!-\! \eta_{O_{k-1}}^{B_{N}}\right) \!+\!\eta^{B_{N}} \!\!-\! \eta_{O_{n-1}}^{B_{N}}\\
& ~~ =  x_{O_A}^{B_{N}}-\eta_{O_A}^{B_{N}}+\eta^{B_{N}}=\eta^{B_{N}}.
\end{split}
\end{equation}
For a standard (undeformed) FRW picture all the non-triviality of our thick slicing is in Bob$_N$'s description of the point of emission $\left( \eta^{B_N}_{O_A},x^{B_N}_{O_A} \right)$, which in a FRW spacetime is Eq.~(\ref{emissFRW}). From~(\ref{BobnAl-CON}), considering also~(\ref{param}),
one finds that\begin{equation}
\begin{split}
& \eta^{B_N}_{O_A} = x^{B_N}_{O_A} = \!-\! \sum_{k=1}^N e^{\sum_{s=k+1}^N \!\! H_s \zeta_s} \frac{ e^{H_k \zeta_k} \!-\! 1}{H_k} \\ &
\rightarrow \!-\! \sum_{k=1}^N  \frac{\zeta}{a(t_k)} \rightarrow \!-\! \int_{-T}^{0} \frac{dt}{{a(t)}}.
\end{split}
\end{equation}

As a last remark we notice that, as we discussed at the beginning of this section, the translational invariance of de Sitter spacetime allowed us to define our thick slices in such a way that each observer Bob$_n$ describes the particle's motion, in the corresponding $n$-th slice (defined in the interval $\eta_{O_{n-1}} \!\! \leq \eta \leq \! \eta_{O_n} $), through the same de Sitter Hamiltonian, the only difference being the value of the  expansion rate $H_n$. Explicitly, in the $n$-th slice,
\begin{equation*}
{\cal H}_{n}^{B_{n}}=\left(1-H_{n}\eta^{B_{n}}\right)^{2}\left(\Omega_{B_{n}}^{2}-\Pi_{B_{n}}^{2}\right)-m^{2}.
\end{equation*}
Using Eqs.~(\ref{BobnBobN-CON}) we get the Hamiltonian that Bob$_N$ attributes to each $n$-th slice:
\begin{equation*}
\begin{split}
& {\cal H}_{n}^{B_{N}} \!\! + m^{2} \!=\! \left(\Omega_{B_{N}}^{2}-\Pi_{B_{N}}^{2}\right) \times \\ &
\times \! \left(\!\! e^{\sum_{k=n+1}^{N} \!\! H_{k}a_{t}} \!\!-\! H_{n} \!\!\!\!\! \sum_{k=n+1}^{N} \!\!\!\! e^{\sum_{s=k+1}^{N} \!\! H_{s}a_{t}}\frac{e^{H_{k}\zeta} \!\!-\! 1}{H_{k}} \!-\! H_{n}\eta^{B_{N}} \!\!\!\right)^{\!\!\!2}
\end{split}
\end{equation*}
One can see that the slices are patched together in such a way that the Hamiltonian is continuous in the point of juncture of two consecutive slices. Indeed
\begin{equation*}
{\cal H}_{n}^{B_{N}}(\eta^{B_N}_{O_n}) = {\cal H}_{n+1}^{B_{N}}(\eta^{B_N}_{O_n}).
\end{equation*}

\section{LIV WITH FRW EXPANSION}
\label{LIVFRW}

The thick-slicing setup introduced in the previous section
is ideally suited for our objective of studying LIV\ and DSR-relativistic effects in spacetimes
with FRW expansion. We start in this section with the LIV case. Through our slicing we will be able to rely on the  results for LIV in de Sitter spacetime reported in sec.~\ref{sec:LIVdeSitter}, thereby obtaining  a LIV-FRW scenario.

In the previous  Sec.~\ref{sec.Slicing} we described FRW expansion in terms of thick slices of de Sitter expansion by also relying on the the translational invariance available in each de Sitter slice.
For the LIV-FRW case we shall of course involve LIV-de Sitter slices, for which, as shown at the end of sec.~\ref{sec:LIVdeSitter}, the breakdown of translational invariance is such that distant observers, at relative rest, describe the motion of particles through different laws, i.e. different functional expressions for the Hamiltonian and velocities (see Eq.~(\ref{velocityLIVdeSitter-2})).

Just as in Sec.~\ref{sec:LIVdeSitter}, we want to find the time of arrival of a soft and a hard photon emitted simultaneously at a distant source. As in the previous section, we take Alice to be local to the event of emission, and Bob$_N$ to be local to the detector. We assume that the soft photon has been emitted at the (comoving) Bob$_N$ time $-T$, and that Bob$_N$'s origin coincides with the event of detection of the soft photon. Moreover we take the intermediate observers Bob$_n$ such that the soft photon crosses the origin of their reference frame. Since in the LIV scenario the laws of transformation among observers  are  not deformed, the relation between Bob$_n$ and Bob$_N$ coordinates is still given by Eqs.~(\ref{BobnBobN-CON}).
We start by assuming that Bob$_N$ describes, in the $N$-th slice, the velocity to be the one given by the Hamiltonian (\ref{dispLIVrep}), Eq.~(\ref{velocityLIVdeSitter}). In order to reconstruct the velocity in the other slices, we can focus on the  scale factor in conformal-time coordinates
\begin{equation}
a_{n}\left(\eta\right)=\frac{1}{1-H_{n}\eta}
\label{an}
\end{equation}
so that (as in Sec.~\ref{sec:LIVdeSitter} we assume $\tilde{\alpha}\!+\!\tilde{\beta}\!=\!1$)
\begin{equation*}
v_{N}^{B_{\!N}} \!\! \left(\eta^{B_{\!N}} \!\right) \!=\! 1 - \Pi^{B_{\!N}} \!\! \left( \!\!\lambda'a_{\!N} \!\left(\eta^{B_{\!N}} \!\right) \!+\! \lambda'' \!\!\!+\! \frac{\lambda}{a_{\!N} \! \left(\eta^{B_{\!N}} \!\right)}+\!\frac{\lambda'''}{a_{\!N}^{2} \!\left(\eta^{B_{\!N}} \!\right)} \! \right)
\end{equation*}
Let us consider first the observer Bob$_{N-1}$. Using relations~(\ref{BobnBobN-CON}) to get the relations between Bob$_{N}$ and Bob$_{N-1}$ coordinates,
\begin{equation*}
\begin{split}
& \eta^{B_{N}}\left(\eta^{B_{N-1}}\right)
=e^{H_{N}\zeta}\left(\eta^{B_{N\!-\!1}}+\eta_{O_{N-1}}^{B_{N}}\right) \\ &
~~~~ = e^{H_{N}\zeta}\left(\eta^{B_{N\!-\!1}}-\frac{1-e^{-H_{N}\zeta}}{H_{N}}\right),
\end{split}
\end{equation*}
we derive Bob$_{N-1}$'s description of the scale factor $a_{N}$ in the $N$-th slice as
\begin{equation*}
a_{N}\left(\eta^{B_{N}}\left(\eta^{B_{N\!-\!1}}\right)\right)
=  \frac{e^{-H_{N}\zeta}}{1-H_{N}\eta^{B_{N\!-\!1}}}=e^{-H_{N}\zeta}a_{N}\left(\eta^{B_{N\!-\!1}}\right)
\end{equation*}
In its origin, since $a_{N}\left(0\right)=1$, Bob$_{N-1}$ describes the scale factor to be $e^{-H_{N}\zeta}$.
We define the $N\!-\!1$-th slice to be such that the scale factor remains constant at the value it has in Bob$_{N-1}$'s origin, the point of contact between the two slices.
This condition can be imposed assuming that in the $N-1$-th slice, governed by the constant expansion rate $H_{N-1}$, Bob$_{N-1}$ describes the scale factor to be
\begin{equation}
e^{-H_{N}\zeta}a_{N-1}\left(\eta^{B_{N-1}}\right).
\end{equation}
With this condition, Bob$_N$ describes the scale factor to be, in the
$N \!-\!1$-th slice,
\begin{equation*}
 e^{-H_{\!N}\zeta}a_{\!N\!-\!1} \!\left(\eta^{B_{\!N\!-\!1}} \!\!\left(\eta^{B_{\!N}}\!\right) \! \right)
 \!=\! e^{-H_{\!N}\zeta}a_{\!N\!-\!1} \!\left(\!e^{\!-H_{\!N}\zeta} \!\!\left( \!\eta^{B_{\!N}} \!\!-\! \eta_{O_{\!N\!-\!1}}^{B_{\!N}} \!\right) \!\!\right)
\end{equation*}
Iterating this construction one finds that Bob$_N$ describes the scale factor  in the $n$-th slice to be
\begin{equation*}
\begin{split}
& a_{n}^{B_{\!N}} \!\!\left(\eta^{B_{\!N}} \!\right) \!\!=\! e^{-\!\!\sum_{k=n+1}^{N} \!\!H_{k}\zeta} a_{n} \! \left(\! e^{-\!\!\sum_{k=n+1}^{N} \!\! H_{k}\zeta} \!\left( \! \eta^{B_{\!N}} \!\!\! -\! \eta_{O_n}^{B_{\!N}} \!\right) \!\!\right)\\ &
~~~~~~~ = \frac{e^{-\sum_{k=n+1}^{N}H_{k}\zeta}}{1-H_{n}e^{-\sum_{k=n+1}^{N}H_{k}\zeta}\left(\eta^{B_{N}}-\eta_{O_n}^{B_{N}}\right)}.
\end{split}
\label{eq:anBN}
\end{equation*}

Consistently with this setup we have that the velocity assigned by
Bob$_N$ to the photons in each $n$-th slice is
\begin{equation}
v_{n}^{\!B_{\!N}}  \!\!\!=\!\! 1 \!-\! \Pi^{B_{N}} \!\!\!\left(\!\! \lambda' \!a_{n}^{\!B_{\!N}} \!\!\left( \!\eta^{B_{\!N}} \!\right) \!+\!\! \lambda'' \!\!\!+\! \frac{\lambda}{\!a_{n}^{\!B_{\!N}} \!\!\left(\eta^{B_{\!N}}\! \right)\!} \!+\! \frac{\lambda'''}{\!\!\left( \! a_{n}^{\!B_{\!N}} \!\! \left(\eta^{B_{\!N}} \!\right)\!\! \right)^{\!\!2} \!} \!\!\right)
\label{velBN-LIV-N}
\end{equation}
Notice that in this way the velocity (and then the Hamiltonian) is continuous in the point of junction of two contiguous slices.
The fact that, as discussed in Sec.~{\ref{sec:LIVdeSitter}}, the term proportional to $\lambda$ in~(\ref{dispLIVrep}) plays a special role in relation to  translational invariance in the de Sitter case, still has a role in the FRW case, even though time-translations are not a symmetry of the FRW case.
We shall see this by first observing that, as a result of (\ref{velBNBn}),(\ref{BobnBobN-CON}) and~(\ref{an}), each observer Bob$_n$ describes the velocity, in the $n$-th slice, to be
\begin{equation*}
\begin{split}
& v_{n}^{B_{n}}\left(\eta^{B_{n}},\Pi^{B_{n}}\right)=  v_{n}^{B_{N}}\left(\eta^{B_{N}}\left(\eta^{B_{n}}\right),\Pi^{B_{N}}\left(\Pi^{B_{n}}\right)\right)\\
& =  1-\Pi^{B_{n}}\bigg(\lambda'e^{-2\sum_{k=n+1}^{N}H_{k}\zeta}a_{n}\left(\eta^{B_{n}}\right)+\lambda''e^{-\sum_{k=n+1}^{N}H_{k}\zeta} \\&
~~~~~~~~~~~~~~~~~~~~~~~~~~~~~~~~ +\frac{\lambda}{a_{n}\left(\eta^{B_{n}}\right)}+\lambda'''\frac{e^{\sum_{k=n+1}^{N}H_{k}\zeta}}{a_{n}^{2}\left(\eta^{B_{n}}\right)}\bigg)
\end{split}
\end{equation*}
One can see that for the term proportional to $\lambda$  the velocity  takes the same form in all slices (though of course with a different expansion rate $H_n$).

In order to evaluate the arrival time at the detector, we consider that the trajectory that Bob$_N$ assigns to photons in the $N$-th slice is
\begin{equation}
x^{B_{N}} \! \left(\eta^{B_{N}} \!\right)_{N} \!\! =\! x_{O_A}^{B_{N}} \!+\!\!\! \sum_{n=1}^{N-1} \!\! \int_{\eta_{O_{n-1}}^{B_{N}}}^{\eta_{O_n}^{B_{N}}} \!\!\!\!\!\! v_{n}^{B_{N}}d\eta^{B_{N}} \!+\!\! \int_{\eta_{O_{N-1}}^{B_{N}}}^{\eta^{B_{N}}} \!\!\!\!\!\!\!\!\!\!\!\! v_{N}^{B_{N}}d\eta^{B_{N}}
\label{trajectoryBobN-LIV}
\end{equation}
The expression~(\ref{velBN-LIV-N}) can be easily integrated in each slice.

The intermediate observers Bob$_n$ are along the worldline of the soft photon, but the hard photon
crosses the spatial origin of these intermediate observers at $\eta^{B_n} \neq 0$,
\begin{equation}
\begin{gathered}
\eta^{B_N}_{O_n}=\eta^{B_N}(\eta^{B_n}=0,x^{B_n}=0)+O(\lambda),\\
x^{B_N}_{O_n}=x^{B_N}(\eta^{B_n}=0,x^{B_n}=0)+O(\lambda).
\end{gathered}
\label{XBN-orBn}
\end{equation}
However in Eq.~(\ref{trajectoryBobN-LIV}) the coordinates $\eta^{B_N}_{O_n},x^{B_N}_{O_n}$ appear only multiplied by factors of $\lambda$ and therefore we can neglect higher orders and obtain from~(\ref{BobnBobN-CON}) the relations
\begin{gather}
\eta_{O_n}^{B_{N}} = -\sum_{k=n+1}^{N}e^{\sum_{s=k}^{N}H_{s}\zeta}\frac{1-e^{-H_{k}\zeta}}{H_{k}}, \label{O-1}\\
\eta_{O_n}^{B_{N}}-\eta_{O_{n-1}}^{B_{N}} = e^{\sum_{s=n}^{N}H_{s}\zeta}\frac{1-e^{-H_{n}\zeta}}{H_{n}}, \label{O-2}\\
\eta_{O_{n-1}}^{B_{N}}+\eta_{O_n}^{B_{N}}=\eta_{O_{n-1}}^{B_{N}}-\eta_{O_n}^{B_{N}}+2\eta_{O_n}^{B_{N}}, \label{O-3}
\end{gather}
Using also that $x_{O_A}^{B_{N}}=\eta_{O_A}^{B_{N}}$
one gets, after some algebra, the trajectory
\begin{equation}
\begin{split}
& x^{B_{N}}\left(\eta^{B_{N}}\right)_{N} = \eta^{B_{N}}-\Pi^{B_{N}}\sum_{n=1}^{N}\Bigg[\lambda'\zeta+\lambda''e^{\sum_{s=n}^{N}H_{s}\zeta}\frac{1-e^{-H_{n}\zeta}}{H_{n}}\\
 & +\lambda e^{2\sum_{s=n}^{N}H_{s}\zeta}\frac{1-e^{-2H_{n}\zeta}}{2H_{n}}+\lambda'''e^{3\sum_{s=n}^{N}H_{s}\zeta}\frac{1-e^{-3H_{n}\zeta}}{3H_{n}}\Bigg]\\
 & -\Pi^{B_{N}}\Bigg[\lambda'\frac{1}{H_{N}}\ln\left[1+H_{n}e^{-\sum_{k=n+1}^{N}H_{k}\zeta}\left(\eta_{O_n}^{B_{N}}-\eta^{B_{N}}\right)\right] \\ & +\left(\lambda''+\lambda+\lambda'''\right)\eta^{B_{N}}\\
 & -\left(\frac{\lambda}{2}+\lambda'''\right)H_{N}\left(\eta^{B_{N}}\right)^{2}+\frac{\lambda'''}{3}H_{N}^{2}\left(\eta^{B_{N}}\right)^{3}\Bigg]
\end{split}
\label{trajectoryBobN-2}
\end{equation}
The time of arrival of the hard photon is obtained by solving $\eta^{B_{N}}\left(x^{B_{N}}=0\right)$. Since
the delay is $O\left(\lambda\right)$, we can disregard all the terms
$O\left(\left(\eta^{B_{N}}\right)^{2}\right)$ in the right-hand side of (\ref{trajectoryBobN-2}),
and we get the expression for the delay
\begin{equation}
\begin{split}
& \Delta\eta^{B_{N}} =  \Pi^{B_{N}}\sum_{n=1}^{N}\Bigg[\lambda'\zeta+\lambda''e^{\sum_{s=n}^{N}H_{s}\zeta}\frac{1-e^{-H_{n}\zeta}}{H_{n}}\\
 & +\lambda e^{2\sum_{s=n}^{N}H_{s}\zeta}\frac{1-e^{-2H_{n}\zeta}}{2H_{n}}+\lambda'''e^{3\sum_{s=n}^{N}H_{s}\zeta}\frac{1-e^{-3H_{n}\zeta}}{3H_{n}}\Bigg]
\end{split}
\end{equation}
In the limit in which $N\rightarrow \infty$, using Eqs.~(\ref{Riemann}) and (\ref{Riemann-2}), the delay takes the form (omitting the
suffix $B_{N}$)
\begin{equation}
\Delta t = \Delta \eta = p_h \int_{-T}^{0}dt\left[\lambda'+\frac{\lambda''}{a\left(t\right)}+\frac{\lambda}{a^{2}\left(t\right)}+\frac{\lambda'''}{a^{3}\left(t\right)}\right],
\end{equation}
where again we expressed the delay in terms of comoving time and we denoted by $p_h$ the momentum of the hard particle observed at the detector.
We can finally reexpress the delay in terms of the redshift of the source $z\equiv z(-T)$ noting that for $\bar{z} \equiv z(t) $
\begin{equation}
a\left(t\right)=\frac{1}{1+\bar{z}}, \qquad
dt=-\frac{d\bar{z}}{H\left(\bar{z}\right)\left(1+\bar{z}\right)},
\label{redshift}
\end{equation}
so that the delay becomes
\begin{equation}
\Delta t=p_h\int_{0}^{z}\frac{d\bar{z}}{H\left(\bar{z}\right)}\left[\frac{\lambda'}{\left(1+\bar{z}\right)}+\lambda''+\lambda\left(1+\bar{z}\right)+\lambda'''\left(1+\bar{z}\right)^{2}\right]
\label{gacextratoo}
\end{equation}
For the choice of parameters $\lambda'=\lambda''=\lambda'''=0$, $\lambda\neq 0$, the delay coincides with the one reported in~\cite{jacobpiran}, as it should have been expected on the basis of the observations we offered above.
Indeed one can show that the trajectory~(\ref{trajectoryBobN-2}), in the limit $N\rightarrow\infty$, is consistent with the Hamiltonian
\begin{equation}
{\cal H} \!+\! m^2 \!\!= \!\! \frac{\Omega^2 \!\!-\! \Pi^2 \!\!}{a^2(\eta)} + \frac{\tilde{\alpha} \Omega^3 \!\!+\! \tilde{\beta} \Omega \Pi^2 \!}{a^2(\eta)} \!\! \left( \!\! \lambda' \! a(\eta) \!+\! \lambda'' \!\!\!+\! \frac{\lambda}{\! a(\eta)} \!+\! \frac{\lambda'''\!}{ \!a^2(\eta)} \!\! \right)
\label{dispLIV-FRW}
\end{equation}
which coincides, for the same choice of parameters $\lambda'=\lambda''=\lambda'''=0$, $\lambda\neq 0$, with the one presented in~\cite{jacobpiran}  and ~\cite{PiranMartinez}.

\section{DSR and Relative locality WITH FRW EXPANSION}
Our final task is to perform a DSR-relativistic analysis with FRW expansion.
 We use again as illustrative application an analysis of  the times of arrival of a soft and a hard photon emitted simultaneously at a distant source, with Alice local to the event of emission, and Bob$_N$ local to the event of detection of the soft photon, emitted at the (comoving) Bob$_N$ time $-T$. Evidently for a DSR-relativistic analysis our ``thick slicing" of FRW must involve slices described by the DSR-deformed de Sitter scenario of sec.~\ref{sec.DSR-dS}.
Analogously to sec.~\ref{sec.Slicing}, the translational invariance (under the $\ell$-deformed translations~(\ref{TransCoord-DSR})) of the DSR-de Sitter setup of sec.~\ref{sec.DSR-dS}, allows us to construct our slices choosing the Hamiltonian in the $n$-th slice, for each observer Bob$_n$, to have the same functional expression as the one of sec.~\ref{sec.DSR-dS}, but with the corresponding value of the expansion rate $H_n$. It follows that in the $n$-th slice, Bob$_n$ describes the photons moving with velocity (for $\Pi>0$)
\begin{equation}
v_{n}^{B_{n}}=1-\ell\left(\gamma+\beta\right)\left(1-H_{n}\eta^{B_{n}}\right)\Pi_{B_{n}}.
\end{equation}
Moreover, each observer Bob$_n$ is connected to Alice, who is at the source, by a transformation~(\ref{FiniteTransSlicing}), but with the $\ell$-deformed translation generators given in Eqs.~(\ref{repDSR-DS}),(\ref{PhSpDSRdS}) (with the $H_n$ appropriate for each $n$-th slice). That is Bob$_n$ is connected to Alice by a series of finite $\ell$-deformed space translations (each with the corresponding parameter $\xi_k$) followed by a series of finite $\ell$-deformed time translations (each with the corresponding parameter $\zeta_k,H_k$). Then, the relation between Bob$_N$'s and Alice's coordinates is obtained computing the transformation~(\ref{FiniteTransSlicing}), for $n=N$, where each intermediate step is described by Eq.~(\ref{TransCoord-DSR}), with the relative translation parameter $\zeta_k$ or $\xi_k$ and expansion rate $H_k$. This leads to
\begin{gather}
\eta^{B_N} \!\!= \eta_N^{(0)} \!+\! \ell (\alpha \!-\! \gamma) \sum_{n=1}^N \zeta_n e^{\sum_{k=n}^N H_k \zeta_k} \!\left(\! 1 - H_n \eta_{n-1}^{(0)} \right) \! E_{H_n,n-1}^{(0)} , \nonumber \\
\begin{split}
x^{B_N}  \!\!= x_N^{(0)} -  & \ell (\alpha \!-\! \gamma) \sum_{n=1}^N e^{\sum_{k=n}^N H_k \zeta_k} H_n x_{n-1}^{(0)} E_{H_n,n-1}^{(0)},
\end{split}
\nonumber \\
\begin{split}
\Omega^{B_N} \!\!=\! \Omega_N^{(0)} + \, & \ell (\alpha \!-\! \gamma) \sum_{n=1}^N e^{-\sum_{k=n}^N H_k \zeta_k}  H_n \zeta_n \Omega_n^{(0)} E_{H_n,n-1}^{(0)},
\end{split}
\nonumber\\
\Pi^{B_N} \!\!\!=\! \Pi_N^{(0)} \!\!+\! \ell (\alpha \!-\! \gamma) \!\sum_{n=1}^N \!e^{-\!\sum_{k=n}^N \! H_k \zeta_k} H_n \zeta_n \Pi^{(0)}_{n-1} E_{H_n,n-1}^{(0)},
\label{TransCoord-N}
\end{gather}
where
\begin{equation}
E_{H_n,k}^{(0)}=\Omega_k^{(0)}\left( 1-H_n\eta_k^{(0)} \right)+H_n\Pi_k^{(0)}x_k^{(0)},
\label{TransCoord-N-2}
\end{equation}
and
\begin{equation}
\begin{gathered}
\eta_k^{(0)} = e^{\sum_{s=1}^k H_s \zeta_s} \eta^A + \sum_{s=1}^k e^{\sum_{r=s+1}^k H_r \zeta_r} \frac{1-e^{H_s \zeta_s}}{H_s} , \\
x_k^{(0)} = e^{\sum_{s=1}^k H_s \zeta_s} \left( x^A - \sum_{s=1}^N \xi_s \right) , \\
\Omega_k^{(0)} = e^{-\sum_{s=1}^k H_s \zeta_s}\Omega^A , ~~~
\Pi_k^{(0)} = e^{-\sum_{s=1}^k H_s \zeta_s}\Pi^A .
\end{gathered}
\label{TransCoord-N-3}
\end{equation}
From these relations we see that, as in Sec.~\ref{sec.DSR-dS}, relative locality affects the coordinates that Bob$_N$ assigns to the point of emission $\eta^{B_N}_{O_A},x^{B_N}_{O_A}$, which now depend on the particles momenta.

We impose again the conditions~(\ref{param}):
\begin{equation}
 \zeta_n = \zeta = T/N , \qquad
 \xi_n = e^{-\sum_{k=1}^{n}H_{k} \zeta_n} \frac{e^{H_{n} \zeta_n} - 1}{H_{n}},
 \label{param-2}
\end{equation}
This amounts to enforcing that every observer Bob$_n$ is local, at the time $\eta(t_n)$, to the trajectory of a soft photon, for which the effects of the deformation can be neglected, emitted at Alice.

To evaluate the time of arrival of the hard photon, we consider the trajectory that Bob$_N$ assigns to photons in the $N$-th slice, which is given by
\begin{equation*}
x^{B_{N}} \! \left(\eta_{B_{N}} \!\right)_{N} \!\! =\! x_{O_A}^{B_{N}} \!+\!\!\! \sum_{n=1}^{N-1} \!\! \int_{\eta_{O_{n-1}}^{B_{N}}}^{\eta_{O_n}^{B_{N}}} \!\!\!\!\!\! v_{n}^{B_{N}}d\eta \!+\!\! \int_{\eta_{O_{N-1}}^{B_{N}}}^{\eta^{B_{N}}} \!\!\!\!\!\!\!\!\! v_{N}^{B_{N}}d\eta .
\label{worldlineBobN}
\end{equation*}
It is straightforward to verify, using Eqs.~(\ref{TransCoord-N}) to establish the relation between the coordinates of  Bob$_N$ and the coordinates of Bob$_n$, that the velocity of the photons  in the $n$-th slice is described by Bob$_N$ by simply re-expressing $v^{B_n}_n$ in terms of the coordinates of Bob$_N$:
\begin{equation}
v^{B_N}_n(X^{B_N}) = v^{B_n}_n(X^{B_n}(X^{B_N})),
\end{equation}
so that
\begin{equation}
\begin{split}
& v_{n}^{B_{N}}(\eta^{B_N}) =  1-\ell\left(\gamma+\beta\right)\Bigg(e^{\sum_{s=n+1}^{N}H_{S}\zeta} \\ &
- H_{n}\sum_{k=n+1}^{N}e^{\sum_{s=k}^{N}H_{S}\zeta}\frac{1-e^{-H_{k}\zeta}}{H_{k}}-H_{n}\eta^{B_{N}}\Bigg)\Pi_{B_{N}}
\end{split}
\label{velBNBn-DSR}
\end{equation}

Eq.~(\ref{velBNBn-DSR}) can be easily integrated in each $n$-th slice, and, after some algebra, the trajectory~(\ref{worldlineBobN}) can be rearranged as follows
\begin{equation}
\begin{split}
& x^{B_N} \!\!=\! x_{O_A}^{B_{N}} \!\!-\! \eta_{O_A}^{B_{N}} \!\!+\! \eta^{B_{N}} \!\!-\! \ell\left(\!\gamma \!+\! \beta\right)\Pi_{B_{N}} \!\left( \!\eta^{B_{N}} \!\!-\! \frac{H_{n}}{2} \!\left(\eta^{B_{N}} \!\right)^{2}\right)\\
 & +\ell\left(\gamma+\beta\right)\Pi_{B_{N}}\sum_{n=1}^{N}\left(\eta_{O_{n-1}}^{B_{N}}-\eta_{O_n}^{B_{N}}\right) \bigg[e^{\sum_{k=n+1}^{N}H_{k}a_{t}}-\\&
 H_{n}\sum_{k=n+1}^{N}e^{\sum_{s=k}^{N}H_{s}a_{t}}\frac{1-e^{-H_{k}a_{t}}}{H_{k}}-\frac{H_{n}}{2}\left(\eta_{O_{n-1}}^{B_{N}}+\eta_{O_n}^{B_{N}}\right)\bigg]
\end{split}
\label{trajectoryBobN-DSR}
\end{equation}
Again, the time of arrival of the hard photon is obtained by solving $\eta^{B_N}(x^B_N=0)$. We can identify two contributions to the delay: one comes from the terms proportional to $\ell(\gamma +\beta)$ (second and third raw of Eq.~(\ref{trajectoryBobN-DSR})); the other comes from the term $\eta_{O_A}^{B_{N}}-x_{O_A}^{B_{N}}$, which, as we pointed out above, contains a non-vanishing contribution proportional to $\ell (\alpha - \gamma)$, as one can see by using Eqs.~(\ref{TransCoord-N}) (recall that $\eta^{B_N}_{O_A} \!\!\!=\! \eta^{B_N}(\eta^{A} \!\!=\! 0, x^{A} \!\!=\! 0) $ and $x^{B_N}_{O_A} \!\!\!=\! x^{B_N}(\eta^{A} \!\!=\! 0, x^{A} \!\!=\! 0) $). Thus
\begin{equation}
\Delta \eta = \ell (\gamma +\beta) \Delta \eta_{\gamma+\beta} + \ell (\alpha - \gamma) \Delta \eta_{\alpha - \gamma}
\end{equation}
Considering that at 0-th order in $\ell$ the relations (\ref{O-1}), (\ref{O-2}), (\ref{O-3}) hold, we get
\begin{equation}
\Delta \eta_{\gamma+\beta} = p_h \sum_{n=1}^{N} e^{2\sum_{s=n}^{N}H_{s}\zeta}\frac{1-e^{-2H_{n}\zeta}}{2H_{n}},
\end{equation}
where we denoted again with $p_h$ the momentum of the hard particle observed at the detector.
From relations (\ref{TransCoord-N}), (\ref{TransCoord-N-2}), (\ref{TransCoord-N-3}), (\ref{param-2}), one gets (after a tedious but straightforward derivation)
\begin{equation}
\begin{split}
& \Delta \eta_{\alpha - \gamma} = p_h \sum_{n=1}^{N}\zeta_{n}e^{2\sum_{k=n}^{N}H_{k}\zeta} \\&
\times \left(1-H_{n}\sum_{k=n}^{N}e^{-\sum_{r=n}^{k}H_{r}\zeta}\frac{e^{H_{k}\zeta}-1}{H_{k}}\right)^{2}.
\end{split}
\end{equation}
Using again the relations~(\ref{Riemann}), (\ref{Riemann-2}) and (\ref{DtDeta}) we get that in the limit $N\rightarrow \infty$, the delay is
\begin{equation*}
\begin{split}
\Delta t  = ~& \ell p_h \Bigg( \ell (\beta+\gamma) \int_{-T}^{0}\frac{dt}{a^2(t)}  \\&
+ (\alpha-\gamma) \int_{-T}^{0}dt\left(\frac{1}{a\left(t\right)}+H\left(t\right)\int_{0}^{t}\frac{dt'}{a\left(t'\right)}\right)^{2} \Bigg).
\end{split}
\end{equation*}
Using relations~(\ref{redshift}), the delay can be written in terms of the redshift of the source as
\begin{equation}
\begin{split}
& \Delta t \!=\! \ell p_h \Bigg( \! (\beta+\gamma) \int_{0}^{z}\frac{d\bar{z}\left(1+\bar{z}\right)}{H\left(\bar{z}\right)} \\ &
~~~~ + \!  (\alpha-\gamma) \int_{0}^{z} \frac{d\bar{z}}{
\left( 1 \!+\! \bar{z}\right) \! H \! \left(\bar{z}\right)} \! \left( \! 1 \!+\! \bar{z} \!-\! H  \! \left(\bar{z}\right) \!\! \int_{0}^{
\bar{z}\left(t\right)} \!\!\!\!\! \frac{d\bar{z}'}{H\left(\bar{z}'\right)}\right)^{\!\!2} \Bigg)
\end{split}
\label{gsacextra}
\end{equation}

One of the interesting applications of this DSR-relativistic result (\ref{gsacextra})
is to compare it to the corresponding result (\ref{gacextratoo}) obtained in the LIV case.
This comparison shows that, as expected, in general DSR scenarios and LIV scenarios produce
completely different results, even when applied to the same class of modified dispersion relations.
It is also noteworthy however that there is special case where the two pictures give the same result:
by fixing $\alpha = \gamma$ in the DSR case of (\ref{gsacextra}) one gets the same  formula
obtained by fixing $\lambda' = \lambda''= \lambda'''=0$
in the LIV case of (\ref{gacextratoo}). This also provides a conceptual perspective on our results:
on the DSR side the choice $\alpha = \gamma$ is such that translation transformations are unaffected
(see Eq.~(\ref{algebraDSR-DS}))
by the quantum-gravity scenario just like on the LIV side the 
choice $\lambda' = \lambda''= \lambda'''=0$
is such that translational invariance is unaffected (see Eq.~(\ref{joclambda}))
by the quantum-gravity scenario.
So both in the LIV case and in the DSR case the formula advocated by Jacob and Piran~\cite{jacobpiran} 
$$\Delta t \propto  \int_{0}^{z}\frac{d\bar{z}\left(1+\bar{z}\right)}{H\left(\bar{z}\right)} $$
is applicable only if the quantum-gravity scenario has no implications
for translational invariance and translation transformations.
When the quantum-gravity scenario does have implications
for translational invariance and/or translation transformations
the Jacob-Piran {\it ansatz} does not apply and the DSR scenario gives results different
from the LIV scenario.

\section{Conclusions}
Our research work was motivated by the fact that there are at this point literally hundreds
of publications on applications of Planck-scale-modified dispersion relations in contexts
 involving FRW expansion,
but all these studies have an exclusively heuristic basis. Over the last decade
significant progress was made on the understanding of 
Planck-scale-modified dispersion relations in the flat-spacetime limit
and we now do have a partial but satisfactory understanding of the generalization
to the case of expansion at constant rate, but for FRW expansion before our investigations
we only had heuristic analyses
and very limited understanding of the conceptual issues at stake.

We feel we here provided a significant step forward toward raising the standards
of this phenomenology. Of particular significance is the understanding that the
much-used Jacob-Piran {\it ansatz} implicitly assumes that spacetime translations
are unaffected by the quantum-gravity scenario, and therefore that {\it ansatz} is applicable
exclusively to a corresponding subset of possible quantum-gravity scenarios.
We obtained predictions for both the LIV case and the DSR case applicable
when instead quantum gravity affects spacetime translations, thereby finally
providing a target for those interested in testing Planck-scale modifications
of the dispersion relation in full generality.

For some of our results an important role was played by our setup describing FRW expansion as a series
of stages of de-Sitter expansion, with appropriate conditions for gluing the different ``thick slices".
This provided a safe path for generalizing to the case of FRW expansion the results recently 
obtained for de-Sitter expansion. While quantum-gravity research is certainly full of surprises,
we cannot imagine any quantum-gravity picture that would prevent one from describing
FRW expansion as a series of de-Sitter expansions, and indeed to our knowledge there
is no quantum-gravity result in the literature that would suggest this could be prevented.
Since we have here shown that this ``thick-slicing setup" can play a pivotal role
in phenomenology, of course this issue should attract even more interest in the future. 

\section*{Acknowledgements}
GAC is supported by a grant from the John Templeton Foundation. 
GR is supported by funds provided by the National Science Center under
the agreement DEC-2011/02/A/ST2/00294. AM wish to acknowledge support by the Shanghai Municipality, through the grant No. KBH1512299, and by Fudan University, through the grant No. JJH1512105.

\end{document}